\documentstyle[12pt,epsf,epsfig]{article}
\newfont{\thiplo}{msbm10 scaled\magstep 2}
\newfont{\gothic}{eufb10 scaled\magstep 2}
\newfont{\unc}{eurb10} 
\newskip\humongous \humongous=0pt plus 1000pt minus 1000pt
\def\caja{\mathsurround=0pt}\def\eqalign#1{\,\vcenter{\openup1\jot \caja
        \ialign{\strut \hfil$\displaystyle{##}$&$ 
        \displaystyle{{}##}$\hfil\crcr#1\crcr}}\,}
\newif\ifdtup

\def\eqright #1\cr{\noalign{\hfill$\displaystyle{{}#1}$}}
\def\eqleft #1\cr{\noalign{\noindent$\displaystyle{{}#1}$\hfill}}

\def\oldreffmt#1{\rlap{[#1]} \hbox to 2\parindent{}}

\def\figfmt#1{\rlap{Figure {#1}} \hbox to 1in{}}

\def\sectioneq{\def\theequation{\thesection.\arabic{equation}}{\let
\holdsection=\section\def\section{\setcounter{equation}{0}\holdsection}}}%

\newcounter{holdequation}

\def\begineq #1\endeq{$$ \refstepcounter{equation}\eqalign{#1}\eqno
	(\theequation) $$}
\def\contlimit{\,{\hbox{$\longrightarrow$}\kern-1.8em\lower1ex
\hbox{${\scriptstyle (a\rightarrow0)}$}}\,}
\def\centeron#1#2{{\setbox0=\hbox{#1}\setbox1=\hbox{#2}\ifdim
\wd1>\wd0\kern.5\wd1\kern-.5\wd0\fi
\copy0\kern-.5\wd0\kern-.5\wd1\copy1\ifdim\wd0>\wd1
\kern.5\wd0\kern-.5\wd1\fi}}
\def\centerover#1#2{\centeron{#1}{\setbox0=\hbox{#1}\setbox
1=\hbox{#2}\raise\ht0\hbox{\raise\dp1\hbox{\copy1}}}}
\def\centerunder#1#2{\centeron{#1}{\setbox0=\hbox{#1}\setbox
1=\hbox{#2}\lower\dp0\hbox{\lower\ht1\hbox{\copy1}}}}
\def\lsim{\;\centeron{\raise.35ex\hbox{$<$}}{\lower.65ex\hbox
{$\sim$}}\;}
\def\gsim{\;\centeron{\raise.35ex\hbox{$>$}}{\lower.65ex\hbox
{$\sim$}}\;}

\def\super#1{\ifmmode \hbox{\textsuper{#1}}\else\textsuper{#1}\fi}
\def\textsuper#1{\newcount\holdspacefactor\holdspacefactor=\spacefactor
$^{#1}$\spacefactor=\holdspacefactor}

\def\getcite#1,{\advance\citenumber by1
\def\getcitearg{#1}\def\lastarg{@}
\ifnum\citenumber=1
\ref{#1}\let\next=\getcite\else\ifx\getcitearg\lastarg\let\next=\relax
\else ,\ref{#1}\let\next=\getcite\fi\fi\next}

\def\pom{{\rm P\kern -0.53em\llap I\,}}
\def\spom{{\rm P\kern -0.36em\llap \small I\,}}
\def\sspom{{\rm P\kern -0.33em\llap \footnotesize I\,}}

\relax
\def\contlimit{\,{\hbox{$\longrightarrow$}\kern-1.8em\lower1ex
\hbox{${\scriptstyle (a\rightarrow0)}$}}\,}
\def\upon #1/#2 {{\textstyle{#1\over #2}}}
\relax
\renewcommand{\thefootnote}{\fnsymbol{footnote}}

\sectioneq

\def\til#1{\centeron{\hbox{$#1$}}{\lower 2ex\hbox{$\char'176$}}}
\def\tild#1{\centeron{\hbox{$\,#1$}}{\lower 2.5ex\hbox{$\char'176$}}}
\def\sumtil{\centeron{\hbox{$\displaystyle\sum$}}{\lower
-1.5ex\hbox{$\widetilde{\phantom{xx}}$}}}

\begin{document} 

\begin{titlepage} 

\rightline{\vbox{\halign{&#\hfil\cr
&\today\cr}}} 
\vspace{0.25in} 

\begin{center} 

\centerline{AMS-02, Strongly Self-Interacting Dark Matter, and QUD}

\medskip

Alan R. White\footnote{arw@anl.gov }

\vskip 0.6cm

\centerline{Argonne National Laboratory}
\centerline{9700 South Cass, Il 60439, USA.}
\vspace{0.5cm}

\end{center}

\begin{abstract}  
 The latest AMS-02 electron/positron precision data add to the spectrum knee as direct cosmic ray evidence for an electroweak scale strong interaction. In addition, there is significant evidence for a strong self-interaction of dark matter. QUD is a unique massless SU(5) field theory with an anomaly-generated bound-state S-Matrix that could be an unconventional origin for the Standard Model. The electroweak scale color sextet quark enhanced QCD interaction is the only new physics. Production of multiple vector bosons, that acquire masses via sextet quark pions, will give the AMS positron and electron cross-sections - related vector boson pair production 
 having been seen at the LHC. Stable sextet quark neutrons (neusons) provide a novel form of very strongly interacting dark matter that has the desired experimental properties. Large rapidity hadronic production of neusons will produce the knee.  
\end{abstract} 
 
 \vspace{0.5in}

\renewcommand{\thefootnote}{\arabic{footnote}} \end{titlepage}

\section{Overview}

Extension of the unprecedentedly accurate AMS-02 data\cite{amsr} on positron and electron cross-sections to energies above the electroweak scale has major implications. Before the publication of the data shown in Fig.~1, it was possible to argue\cite{bkw} that
the positron content of cosmic rays is the consequence of galactic collisions of high energy primaries via established physics. Now, it is unarguably clear that a new phenomenon of some kind is necessarily involved. 

\begin{center}  

\epsfxsize=3.8in \epsffile{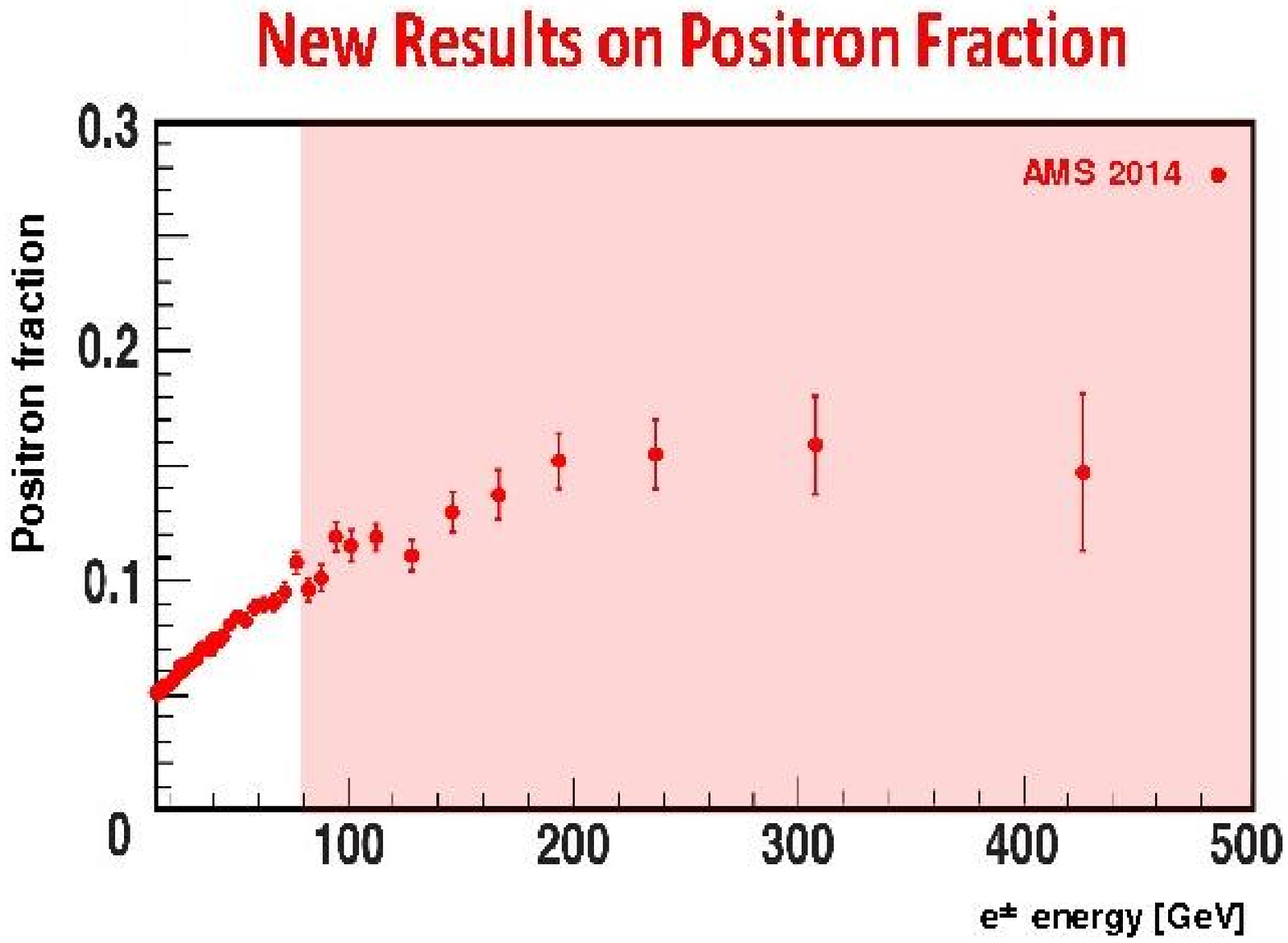}


\epsfxsize=2.7in \epsffile{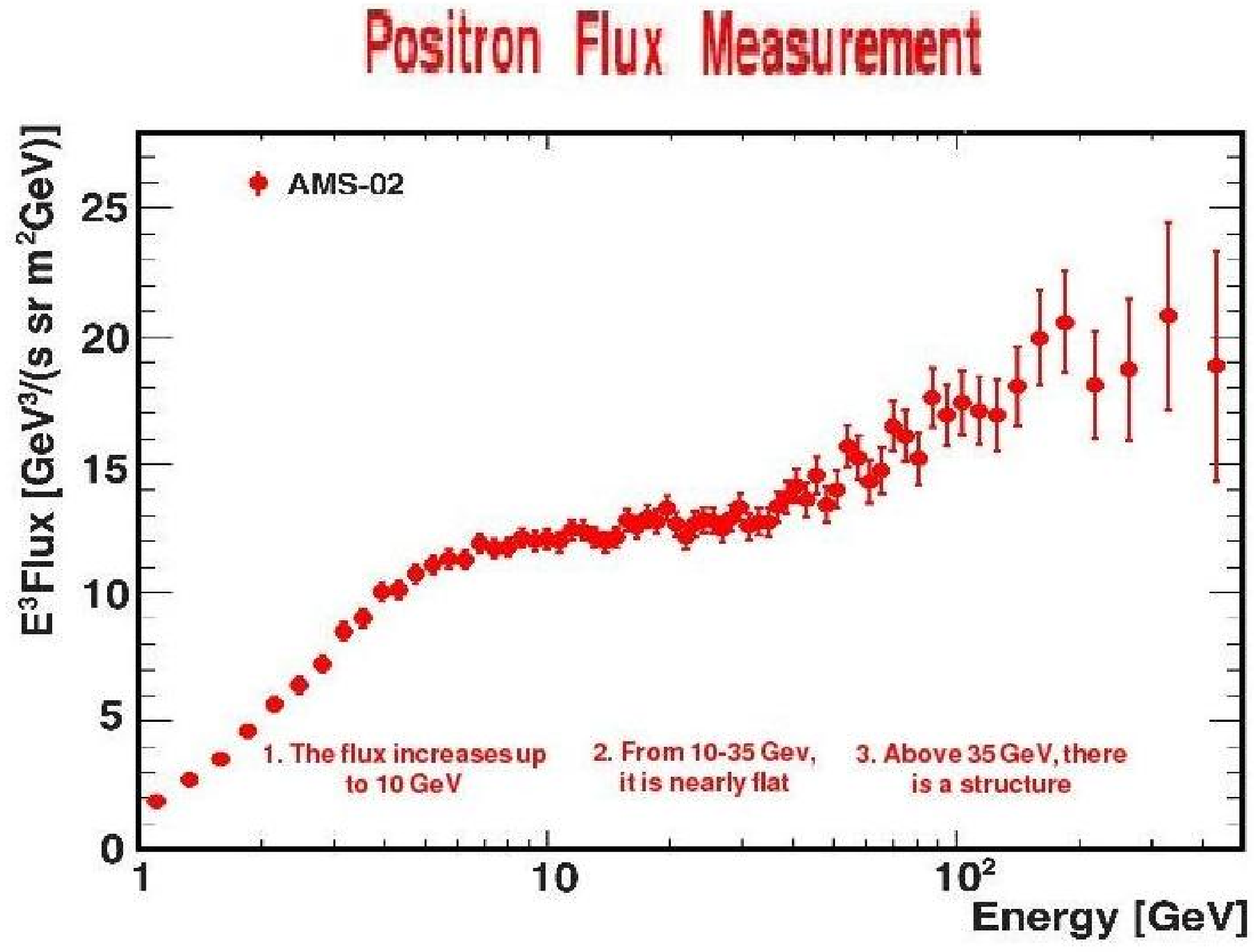} $~~$
\epsfxsize=2.7in \epsffile{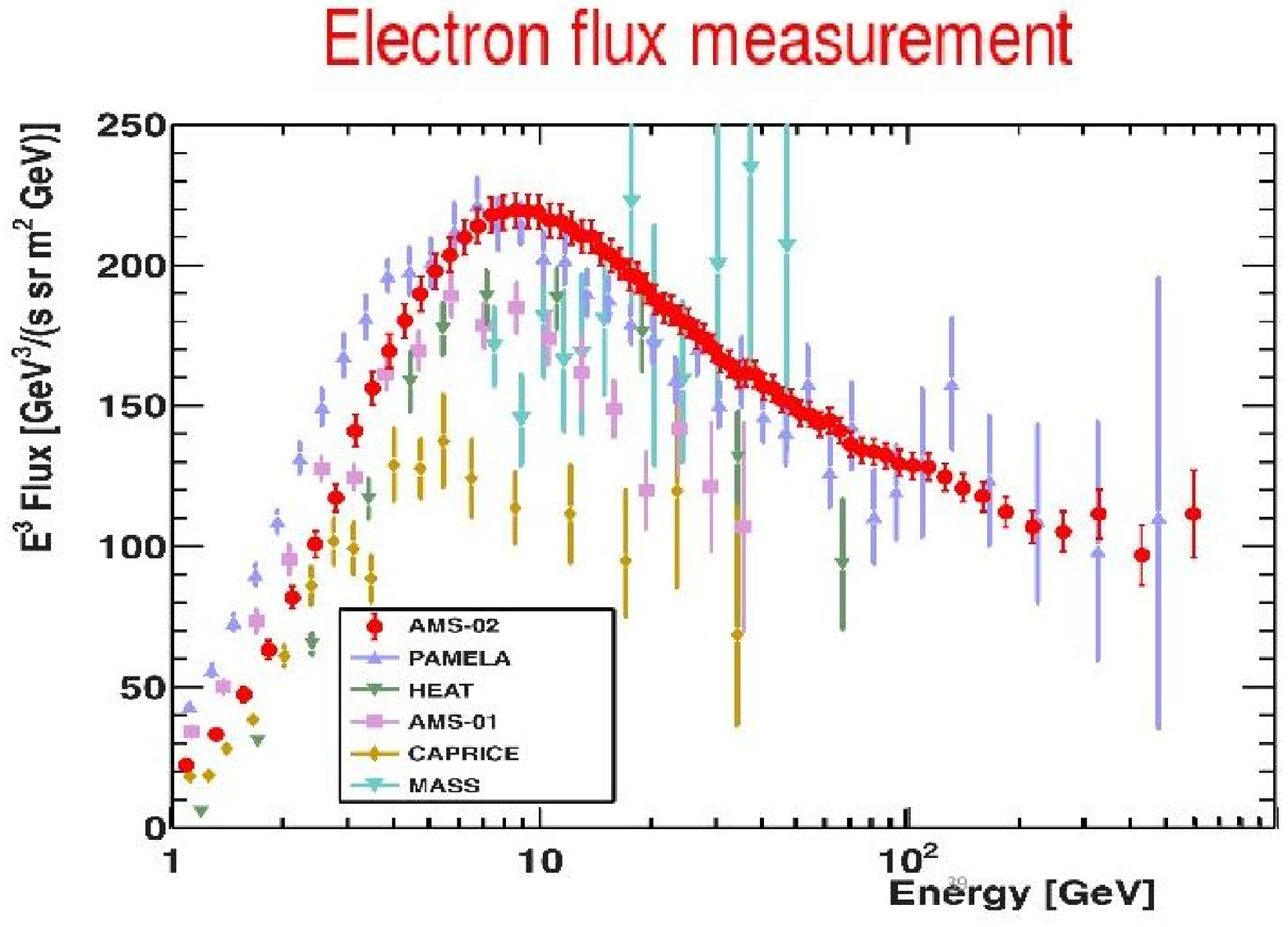}

Figure 1 AMS-02 Data
\end{center}

Various proposals have been put forward but the simplest and most obvious explanation is that there is a new electroweak scale strong interaction, within the primary collisions, that plentifully produces positrons and electrons equally.
This explanation is a radical proposition within the current theory paradigm but the consequences are historic if what is seen is (as I already suggested\cite{arw3} for the lower energy results) the QCD interaction of the color sextet quark sector of the theory\cite{arw1} that I call QUD. Very significantly, the neutral nucleon states of this quark sector  provide a form of dark matter that seems to have just the physical properties now thought to be essential. 

QUD\footnote{Quantum Uno/Unification/Unique/Unitary/Underlying 
Dynamics} is a very special massless field theory that was first discovered 
\cite{kw} 
because of it's unique high-energy unitarity properties. It was then realized that, amazingly, this theory might underly, unify, and provide an origin for, the full Standard Model. It is an infra-red fixed-point SU(5) theory within which novel massless fermion anomaly dynamics lead to a massive bound-state S-Matrix in which the states and interactions are\cite{arw3}-\cite{arw8}, apparently, those of the Standard Model. There is also a Higgs-like boson resonance and so QUD provides an immediate explanation\cite{arw9} for the emergence of the
``nightmare scenario'' at the LHC. 

It would be hard to exaggerate the significance of 
establishing that QUD underlies the Standard Model. 
It is self-contained, can not be added to, involves no explicit parameters, and contains both neutrino masses and dark matter. It has to be entirely right, including the dark matter sector, or else it is simply wrong. So far, there is much apparent success
and no obvious failure. That the multi-regge construction of the S-Matrix is very far outside the mainstream, both conceptually and technologically, may explain, perhaps, why it's virtues remain largely unappreciated.

The most radical paradigm change involved is that     
consistency with the underlying SU(5) unification is achieved, in the  multi-regge (infinite momentum) framework, via the infra-red enhancement of Standard Model interactions by a ``wee parton vacuum'' of anomalous wee gauge bosons coupled to massless fermion anomalies. (The wee gauge bosons are multi-reggeon configurations that
are non-local generalizations of the familiar anomaly current.) As I will outline, the derivation of this crucial result utilises elaborate multi-regge theory. It is in sharp contrast to the conventional wisdom that additional short-distance interactions are needed to achieve
unification of the forces of the Standard Model.  

The color sextet quark sector is the only new QUD physics beyond the Standard Model. It produces both electroweak symmetry breaking and, as already referred to, a very special form of strongly interacting dark matter. The existence of the sextet quark sector is crucial for the unitarity properties of the S-Matrix and, perhaps as a consequence, the dynamical properties of the QCD sector are modified in a manner which improves considerably the comparison with experiment. Particularly
important is the absence of glueballs which, for several years has been a deep puzzle for QCD, but which is vital for QUD dark matter. The Regge pole nature of the pomeron (in first approximation) is also apparent. This has recently been clearly demonstrated\cite{TOT} 
experimentally at the LHC - even though it is commonly believed that it is not a property of QCD.

Because of the large wee parton anomaly color factors involved, sextet QCD interactions are stronger than those of the triplet sector, even though the masses are all electroweak scale. Also, although the anomaly dynamics is subtle, it is the same for the triplet and sextet sectors. Consequently, the sextet matter sector will have many properties analogous to the familiar triplet sector, with the difference being that sextet matter will be more massive, more strongly interacting, and predominantly neutral.
There are two sextet flavors which, therefore, produce sextet ``pions'' and ``nucleons'' as the physical states. The sextet pions become longitudinal vector bosons, while the sextet quark neutrons (neusons) are stable, as I will discuss, and provide dark matter.
 
From the AMS-02 data shown in Fig.~1 we see that, as the electroweak scale is approached (and beyond) there is a dramatic increase in positron production and a corresponding enhancement of electron production. The phenomenon involved is most directly reflected, obviously, in the increase in the positron fraction. If QUD is responsible, then we are looking at ``low-energy'' products of the sextet quark interaction. This interaction will (in parallel with the triplet quark hadronic sector) be dominated by the production of sextet pions that will be manifest, physically, as the production of multiple vector boson states. The leptonic decays of the vector bosons will produce electrons and positrons plentifully, in equal numbers, just as seen by AMS.

Unfortunately, as I have lamented often in previous papers, the kinematic focus of the LHC is completely wrong for the discovery of the QUD strong interaction. Nevertheless, supporting evidence for the QUD explanation of AMS data has now been seen, albeit at
the extreme short-distances studied at the LHC. An excess cross-section for vector boson pair production of the kind predicted by QUD (i.e. $W^+W^-$ and $Z^0Z^0$, but not $W^{\pm}Z^0$, $W^+W^+$, or $W^-W^-$), that increases with energy, has now been measured by both ATLAS\cite{ATWW} and CMS\cite{CMWW}. Nevertheless, it will surely be ironic if it happens that, once again, the discovery of a new interaction is viewed historically as made by cosmic ray physics when it could/should, so easily, have been discovered in an accelerator - certainly the LHC, perhaps even the Tevatron.

Sextet protons (prosons) have an electromagnetic mass and decay into neusons.
Therefore, the neutral neusons can be produced as charged prosons that decay, and so they should appear as ultra-high-energy cosmic rays. It is crucially important, as I have already implied, that there are no glueballs in the QUD S-Matrix. This is because the physical states all have chiral properties associated with anomaly pole components coupled to anomalous wee gluons. Would-be glueballs do not have an anomaly pole component and so can not appear as physical states. Also, because of the absence of hybrids, there are no triplet/sextet mesons
that could provide an interaction between the triplet and sextet sectors. Consequently, the high-energy pomeron provides the only QCD interaction between the two sectors and so dark matter neusons have the fundamental property that they have a strong interaction  with normal matter (i.e. hadrons), but only in very high energy collisions.

As discussed in \cite{arw6}, a pomeron interaction can involve either a rapidity gap or, most often, a high multiplicity hadron state spread uniformly across part of the rapidity axis. The high-energy dominance of sextet states via pomeron interactions, potentially, provides an explanation for the dominance of dark matter production in the ultra-high-energy processes responsible for early universe formation.
It also implies that prosons will be preferentially produced
in the present universe, relative to protons, by the highest energy strong interaction processes. 
Hence, after cosmic acceleration, neusons produced by prosons (and antineusons produced by antiprosons) should be the dominant cosmic rays, at the highest energies.

When first discovered, cold dark matter was conjectured to be very different from
QUD dark matter. Now it seems that more and more of the properties of QUD appear in
the latest models. Composite models are routinely considered\cite{clmx,bfkt} and it is recognized 
that strong self-interactions are essential with cross-sections even bigger than 
those of standard nuclear cross-sections. Both properties are intrinsic for QUD sextet dark matter. This is fortunate since, as we stated above, if QUD is correct then dark matter must be neusons or antineusons (and, perhaps, neuson or antineuson composites). 

Ironically, the most recent dark matter models invoke\cite{bfkt} a complete (hidden sector)  non-abelian gauge theory, in addition to the Standard Model, to produce very strong interactions. 
A priori, it might seem to be much simpler to have a higher mass sector of QCD produce the interaction. However, the conventional expectation is that, because of the evolution of $\alpha_s$, a higher mass sector would necessarily be weaker interacting. This is why a ``hidden sector'', entirely new, gauge theory is invoked. 

Fortunately, the S-Matrix anomaly dynamics of QUD implies that the higher mass QCD sextet sector will, indeed, be more strongly interacting than the lower mass triplet sector. The sextet sector slows down the evolution of $\alpha_S$ beyond the electroweak scale, but this is not the most important phenomenon. More importantly,
the anomalous wee gluon enhancement of the basic interactions is much stronger for 
the higher color sextet sector. As a result, a stronger interacting, more massive sector and a less strongly interacting, lighter sector actually coexist within the same gauge theory. 
 
In \cite{arw6} I discussed a variety of cosmic ray phenomena that can be interpreted as evidence for the QUD sextet quark interaction, including the presence of neusons as a major UHE component that is responsible for all the confusion surrounding the energy dependence of the mass composition. I also discussed why a, large cross-section, electroweak scale interaction is the simple and obvious explanation for a cosmic ray mystery that has been well-known for many years, without any agreement on the phenomena involved. The spectrum knee\footnote{With the selection of data shown in Fig.~2, the ``knee'' looks more like a break in the slope than it does when the full set of (much less precise) data from all experiments is included.} shown in Fig.~2, is widely believed to be astrophysical in origin. 
However, it is such a pronounced (relatively) local effect in the spectrum that it is surely far more plausible that the atmospheric interaction is responsible, with the apparent increase of slope due to the production of undetected particles. That these particles could not be dark matter in any of the forms that in the past were commonly proposed was seen, not surprisingly, as a major argument against such an interaction.
\begin{center}

\epsfxsize=3.5in \epsffile{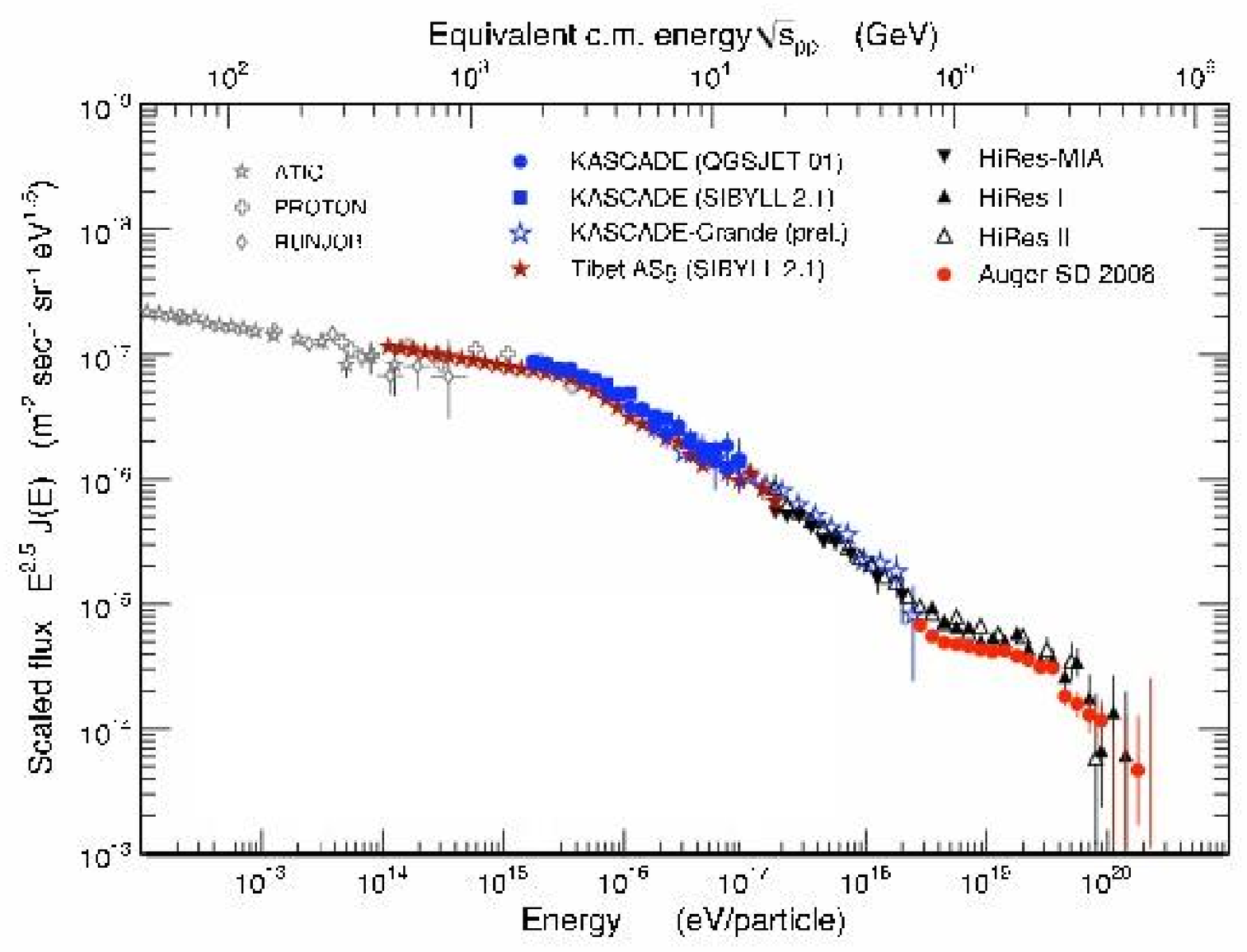}

Figure 2 The Cosmic Ray Spectrum Knee

\end{center}

The QUD sextet quark interaction in the atmosphere can be the origin of the knee, as the threshold for double pomeron production (via either a rapidity gap or a high multiplicity hadron state) of dark matter neusons and antineusons.
As we will discuss, we expect neusons to have a mass $\sim 500$ GeV. In addition, flavor quantum numbers (``sextet isospin'')
require that they be produced multiply - with two being the minimum. Consequently, the energy scale for the knee is, essentially, set by double pomeron production of massive states with a mass $\sim$ 1 TeV, which is very reasonable phenomenologically. There is, however, a minor subtlety. Neusons will be produced both by the atmospheric collisions of normal hadronic primaries and also by the atmospheric collisions of incoming primary neusons.
Because of the large neuson mass the corresponding thresholds will be close in energy\cite{arw6}. In both cases, produced neusons will lose energy via a small
(including zero) number of subsequent collisions and then ``disappear'', thus producing the knee.

\section{QUD and the Origin of the Standard Model}

QUD was initially uncovered\cite{kw,arw10} via a search for a connection between the Critical Pomeron and gauge theory reggeon interactions.
The Reggeon Field Theory Critical Pomeron is the only known description of rising total cross-sections (an essential requirement to match with an
asymptotically-free gauge theory at short distances) that  
satisfies full multiparticle t-channel  unitarity and all s-channel
unitarity constraints. The RFT supercritical phase occurs in superconducting QCD, and 
the critical behavior requires that asymptotic freedom be saturated. The only realistic 
possibility is QCD$_S$, with six color triplet quarks and two color sextet quarks. This  is a real possibility if  ``sextet pions'' are responsible for electroweak symmetry breaking! 

Consequently, the Critical Pomeron requires a specific form of electroweak symmetry breaking that introduces, moreover, a higher color QCD interaction (of the kind seen by AMS!) In addition, a detailed analysis shows that an infra-red fixed-point is also needed and so QCD$_S$ must contain only massless quarks - which is certainly not
physically realistic! At first sight this is fatal for the physical realization of the Critical Pomeron. Fortunately, there is a resolution. The way out is provided by the inclusion of the electroweak interaction. Insisting on both asymptotic freedom and no anomaly, the required sextet quark sector and the electroweak interaction embed 
uniquely\cite{kw} in 
\begin{center}
{\bf QUD $\equiv$ SU(5) gauge theory with left-handed massless fermions in the 
$5 \oplus 15 \oplus40 \oplus 45^*$ representation. Under $ SU(3)\otimes SU(2)\otimes U(1)$}
\end{center}
{\footnotesize $$ 
5=(3,1,-\frac{1}{3}))
+(1,2,\frac{1}{2})~,~~~~~ 15=(1,3,1)+
(3,2,\frac{1}{6}) + (6,1,-\frac{2}{3})~,
$$
$$
40=(1,2,-\frac{3}{2})
+(3,2,\frac{1}{6})+
(3^*,1,-\frac{2}{3})+(3^*,3,-\frac{2}{3}) + 
(6^*,2,\frac{1}{6})+(8,1,1)~,
$$
$$
45^*=(1,2,-\frac{1}{2})+(3^*,1,\frac{1}{3})
+(3^*,3,\frac{1}{3})+(3,1,-\frac{4}{3})+(3,2,\frac{7}{6}))+
 (6,1,\frac{1}{3}) +(8,2,-\frac{1}{2})
 $$}
 
 Strikingly, both the triplet quark and lepton sectors are very close to the Standard Model. Also, QUD is real {\it \{vector-like\}} with respect to $SU(3)_C\otimes U(1)_{em}$. The $SU(2)_L \otimes U(1)$ quantum numbers are not right, but the short-distance lepton anomaly is correct. Having first discovered how the reggeon anomaly dynamics of QCD$_S$ leads to the Critical Pomeron, I realized that, with the same dynamics, QUD could actually be physically realistic. All elementary leptons and quarks would be confined and remain massless, with infra-red anomalies dominating the dynamics! The Standard Model could be an effective theory obtained (in principle) by integrating out the elementary leptons.
  
At first sight, finding the bound-states and interactions of a massless field theory, such as QUD, looks hopeless. Fortunately, multi-regge theory comes to the rescue. The multi-regge region involves multiple infinite-momenta and it is well-known that infinite momentum wee partons can play a vacuum role. In the next Section we outline how QUD infra-red divergences produce wee partons that accompany  multi-regge S-Matrix interactions and states that are, relatively simple and,
 essentially, determined by the parity properties of the fermion couplings. 

The elementary leptons and triplet quarks are sufficient to produce the known states of the Standard Model. Moreover, with an infra-red fixed-point and no exact chiral symmetries, all bound-states will acquire masses via reggeon diagram interactions. 
The sextet sector, which we will be discussing further, is responsible for electroweak symmetry breaking as anticipated. The color octet sector was certainly not asked for. 
However, it plays a crucial role, via large $k_{\perp}$ anomalies, in producing  the generation structure of the physical states. Consequently, the potentially profound possibility is raised that the full array of Standard Model physics could actually be uniquely determined, via QUD, by the necessity to accompany the Critical Pomeron and provide a complete, massive, unitary S-Matrix. 
 
\section{QUD Reggeon Diagram Analysis}

The following is a very brief description of the procedure\cite{arw1,arw6} for constructing the bound-state S-Matrix using multi-regge reggeon diagrams.

Initially, all reggeons are given masses using scalar condensates. Fermion masses are removed first, leaving (fundamentally) chirality transitions that break SU(5) to SU(3)$_C\otimes$U(1)$_{em}~$, but only in reggeon anomaly vertices. Gauge boson masses
are then smoothly removed (via complementarity), producing a sequence of increasing reggeon global symmetries 
 \newline $~$
 \newline
 \centerline{ $\rightarrow SU(2)_C~, \rightarrow  SU(4),~
  \lambda_{\perp} \to \infty,~ \rightarrow SU(5)$}
 
 \noindent The last scalar removed is asymptotically free and so the $k_{\perp}$ cut-off limit $\lambda_{\perp} \to \infty$  can be taken between the SU(4) and SU(5) limits.
 With $\lambda_{\perp} \neq \infty $ many fermion loops violate Ward identities and
all left-handed gauge boson interactions (except those mixing with sextet pions) are eliminated by exponentiating reggeization divergences.
A complete infra-red analysis (using  properties of reggeon interaction kernels) shows that all divergences exponentiate via reggeization, apart from an overall divergence which produces ``universal wee partons'' that select the physical states and interactions, as illustrated in Fig.~3. The wee partons are gauge bosons that form an adjoint SU(5) representation with anomalous color parity ($C\neq \tau$) that keeps them outside the exponentiations of reggeization. Instead, the divergence they produce couples to anomaly vertices that produce anomaly poles via chirality transitions. 

U(1) anomaly poles couple the wee partons in distinct reggeon channels. Chiral Goldstone anomaly poles are a central element of all the bound-states. 
(Goldstone pions are produced within QCD as shown in Fig.~4.) A Goldstone anomaly pole can be characterized either as a bound-state of the intermediate fermion pair (one of which is initially unphysical and carries zero momentum) or, after the chirality transition, as a reggeon state containing wee partons. The reggeon state will have an SU(5) singlet projection, while the bound-state, because of the chirality transition, will violate this symmetry.
 \begin{center}

  \epsfxsize=4.7in \epsffile{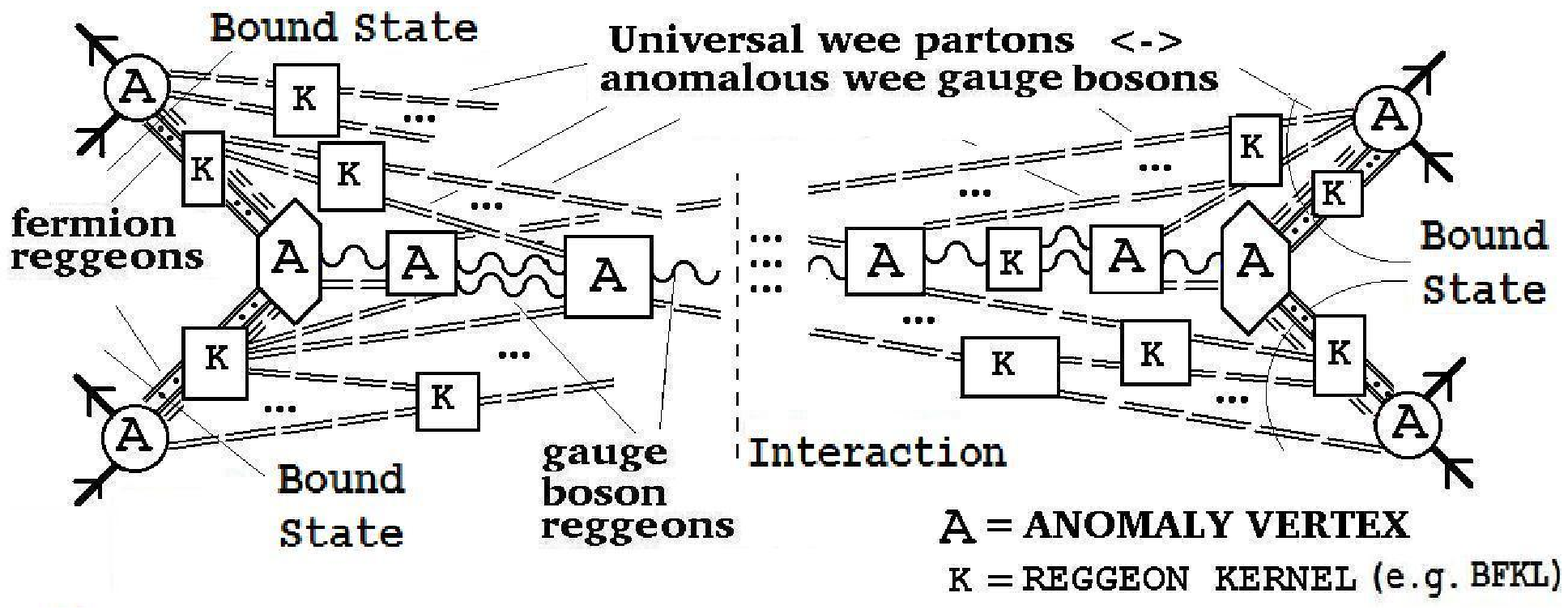}
    
    Figure 3. Reggeon Diagrams Contributing to the Overall Divergence. 
   
 \end{center}
 \begin{center}
\epsfxsize=5in \epsffile{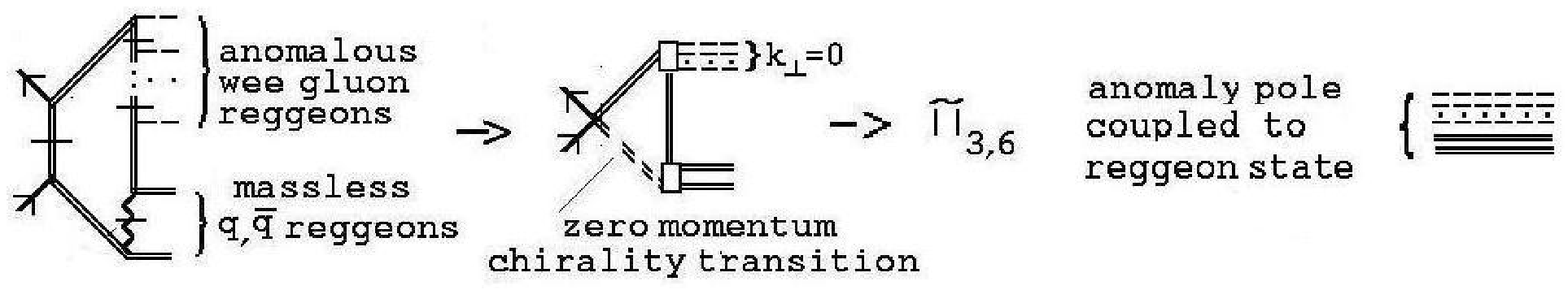}

 Figure 4. The Anomaly Pole Goldstone Pion. 
  \end{center}

 Left-handed gauge bosons contribute as wee gauge bosons only when accompanying the sextet pion / gauge boson interaction. After $\lambda_{\perp} \to \infty$ and the SU(5) limit, the Critical Pomeron and the massless photon appear together. Also, color octet chiral anomalies at $ k_{\perp} = \infty$ produce\cite{arw1} bound-states in Standard Model generations. The resulting ``Standard Model'' vector interactions between bound-states are 
 \begin{itemize}\openup-0.25\jot{
 \item{ The $\tau =  +1$ Critical Pomeron $\approx$ SU(3) gluon reggeon + wee gauge bosons 
 \newline $\leftrightarrow$ SU(5) singlet projection. There is no BFKL pomeron and no odderon.}
 \item{ The $\tau=-1$ Photon $\approx$ a U(1)$_{em}$
 gauge boson + wee gauge bosons 
 \newline $\leftrightarrow$ SU(5) singlet projection.}
 \item{ The Electroweak Interaction $\approx$ left-handed gauge boson mixed with a sextet pion  + wee gauge bosons $\leftrightarrow$ SU(5) singlet projection. The sextet quark right-handed flavor symmetry $\rightarrow$ Standard Model SU(2)$_L$ symmetry.}}
 \end{itemize}
The sextet pions ($\pi_6$) provide a mass for the electroweak vector bosons via anomaly
interactions of the kind illustrated in Fig.~5.
\begin{center}

\epsfxsize=5in \epsffile{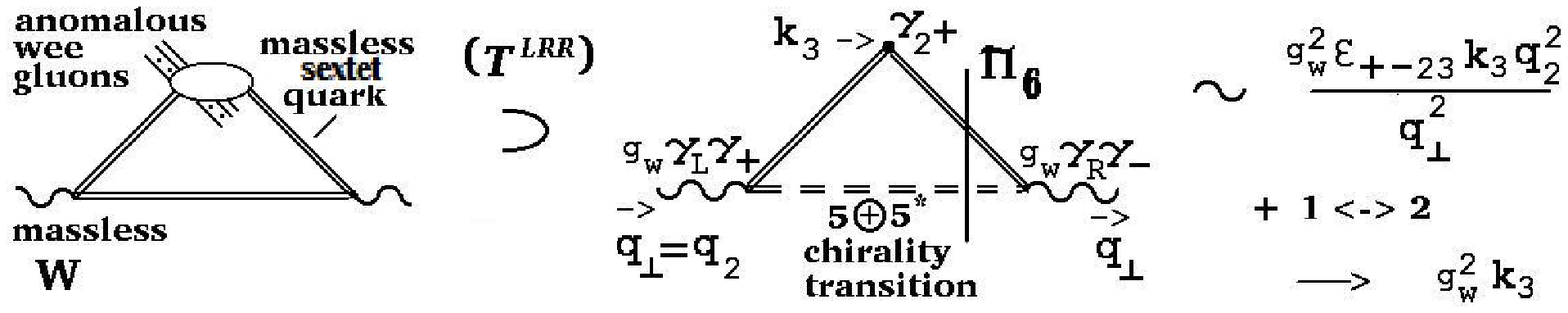}

Figure 5. Vector Boson Mass Generation
\end{center}
This gives a mass $M_W^2$ ($\sim g_W^2 \int dk k$) as a wee gluon integral multiplied by a sextet anomaly factor, while the chirality transition provides a
sextet flavor quantum number that prevents the exponentiation that would result if the
massless vector boson interacted with the wee partons. The perturbative coupling of the $W^{\pm}$ and $Z^0$ is retained and the sextet flavor symmetry becomes the SU(2)$_L$ of the Standard Model.

 The elementary QUD coupling is kept very small by the fixed-point, but physical Standard Model couplings will emerge if the infinite sums of wee gauge boson anomaly color factors enhance couplings sufficiently. The anomaly factors clearly give
$$~\alpha_{\scriptscriptstyle QCD} ~> ~\alpha_{\scriptscriptstyle em}~>>~ \alpha_{\scriptscriptstyle QUD} \sim
 \frac{1}{120}$$
 
 Three Standard Model generations of physical hadrons and leptons appear  
 via the octet anomalies that remain at infinite light-cone momentum after the full SU(5) symmetry is restored. When described in terms of fermions only, without anomalous wee gauge bosons but before the chirality transitions, all bound-states are SU(5) singlets composed of five elementary fermions, two of which are non-dynamical color octets forming  an ``octet pion'' large $k_{\perp}$ anomaly contribution.
 In general, anomaly vertex mixing, combined with fermion and wee parton color factors, will produce a wide range of bound-state mass scales. 
 
In addition to the color octets, lepton bound states contain three elementary leptons with two producing an anomaly pole. The electron is very close to elementary because the anomaly pole involved disturbs the Dirac sea minimally. The muon has the same constituents, but in a different anomaly pole dynamical configuration
that will obviously generate a significant mass. The very small QUD coupling should be the origin of desirably small neutrino masses - that have no color or electric charge anomaly enhancement via wee partons.  Anomaly color factors imply 
    $$M_{\scriptstyle hadrons} >> M_{\scriptstyle leptons} >>  M_{\scriptstyle \nu 's} ~\sim~ \alpha_{\scriptstyle QUD}$$
  
Non-leptonic bound-states are triplet or sextet quark mesons and baryons. 
There are no hybrids and, because the anomaly pole bound-states are necessarily 
chiral Goldstones, there are no glueballs.
Two QUD triplet quark generations give Standard Model hadrons - that mix appropriately. The physical b quark is a mixture of all three QUD generations.
There are two ``exotic'' triplet quarks with charges -4/3 \& 5/3 that have no
chiral symmetry and so do not produce light (anomaly pole) bound-states. The left-handed ``top quark'' forms an 
electroweak doublet with one of the
exotic quarks and does not appear in low mass states. However, a bound-state involving the right-handed top quark remains that 
mixes with the sextet $\eta$ to give two mixed-parity scalars\cite{arw4,arw5} - the $\eta_3$ and the $\eta_6$. The $\eta_6$ has an electroweak scale mass, while the $\eta_3$
has a mass between the triplet and sextet scales. ``Tree-unitarity'' implies that the combined $\eta_3$ and $\eta_6$ couplings should reproduce\cite{arw5} the Standard Model 
electroweak couplings of the Higgs boson. Therefore, the $\eta_3$  could  be the ``Higgs boson'' discovered at the LHC.
If so, the ``QUD Higgs'' is, predominantly a ``top/anti-top'' resonance.   

At first sight, top quark physics is very different from the Standard Model. However, because the sextet ${\eta}$  reproduces the Standard Model final states (automatically at an electroweak scale
mass~!!) it is hard, experimentally, to distinguish the difference\cite{arw3,arw6}. 
A fundamental outcome is that the ``top quark mass'' of the Standard Model is electroweak scale simply because a sextet quark resonance is, essentially, responsible, for the events. An important, experimentally distinguishing, feature of QUD ``top quark physics'' is that the $\eta_6$ should be seen (and may have already been seen\cite{arw3,arw6}) decaying into $Z^0Z^0$ pairs at the LHC - at the experimental 
 $t\bar{t}$ threshold (where $t$ is the Standard Model top quark).
  
\section{The Sextet Scale, Positrons, and Dark Matter}

It is well-known that, based on feynman graph color factors,  
triplet and sextet quark momentum scales 
are expected to satisfy (approximately) the ``Casimir Scaling'' rule  
$$
C(6)~\alpha_s(P_6^2)~\sim ~C(3) ~\alpha_s(P_3^2)
$$
where $P_3$ and $P_6$ are dynamical momenta and $C(3)$ and $C(6)$ are Casimirs for triplet and sextet quarks respectively.
Equivalently, in going from triplet to sextet graphical contributions, $\alpha_s$ is effectively replaced by $\{C(6)/C(3)\}~ \alpha_s$.) 

For SU(3) there are two Casimir operators which, in terms of the generators $G_a$ can be written as 
$$
C_2~=~G^2~ \sim ~f_{abc}G_aG_bG_c ~,~~~~ C_3~\sim~d_{abc}G_aG_bG_c
$$
$C_2$ appears predominantly in anomaly color factors and it is at the origin of the large magnitude of sextet anomaly color factors. Since $C_2(6)/C_2(3)= 5/2$ and $C_3(6)/C_3(3)~=~ 7/2$, we can write
$$
3 ~\alpha_s(P_6^2)~\sim ~\alpha_s(P_3^2)
$$
\begin{center}

\epsfxsize=2.3in \epsffile{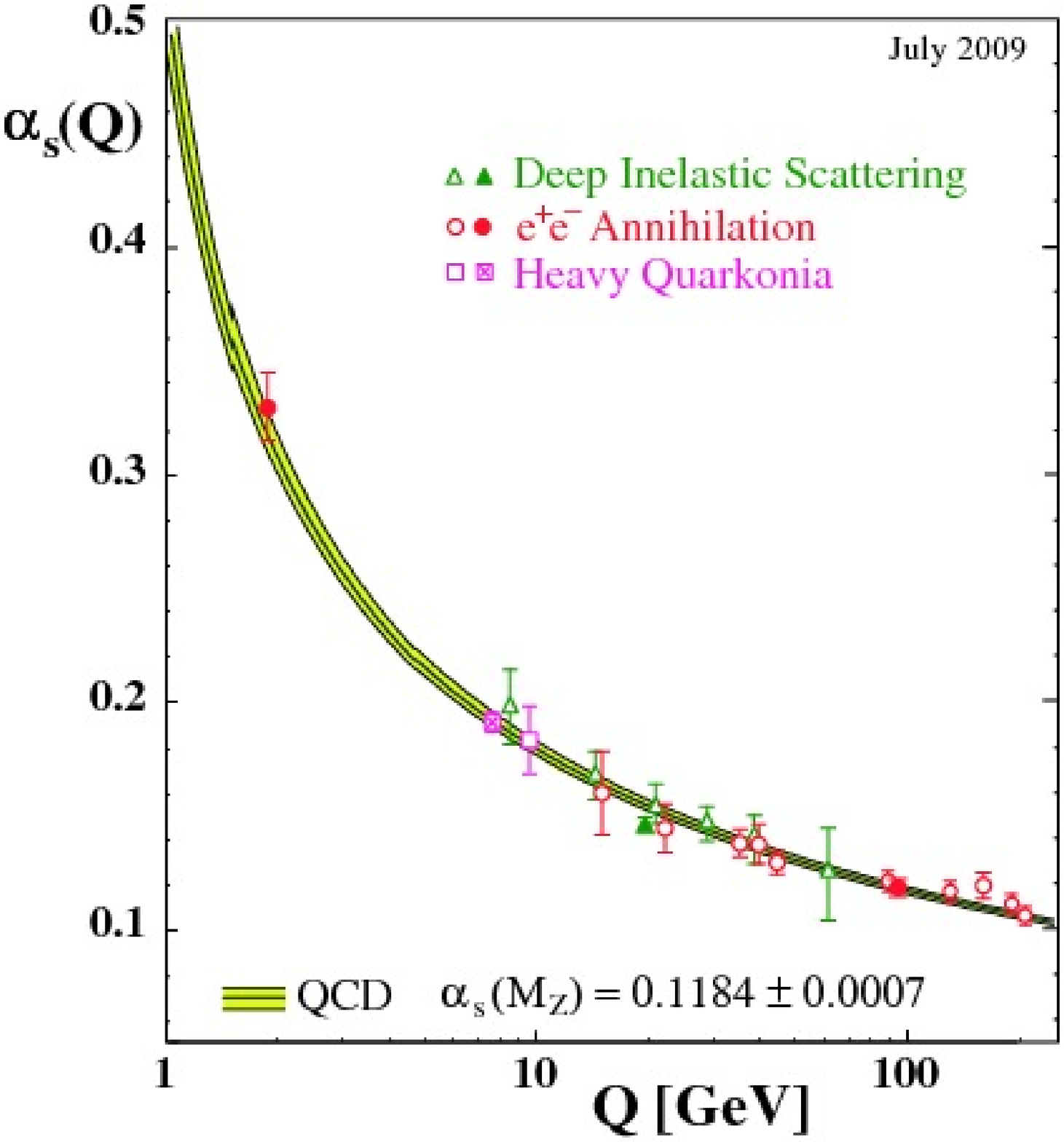}

Figure 6. $\alpha_S(Q)$ 
  \end{center}
From Fig.~6, we can see that $P_3 \sim 1$ GeV corresponds to $P_6 \sim 100$ GeV.
Approximately, therefore, the order of magnitude of the sextet dynamical scale is indeed the electroweak scale! Consequently, 
\begin{itemize}
\item{\it the masses of QUD sextet states will be 
electroweak scale.}
\end{itemize}
\noindent In addition,
\begin{itemize} 
\item{\it wee gluon anomaly color factors will
give large pomeron couplings to sextet states - producing
large high-energy cross-sections.}
\end{itemize}

The presence of the sextet sector affects both the evolution of $\alpha_s$ and the extraction of parton distribution functions. 
In the early studies of jet cross-sections at the Tevatron an infamous ``jet excess'' was discovered that could have (should have?) been viewed as direct evidence for the sextet sector. The results for both $\alpha_s$ and the jet cross-section, shown in Fig.~7, are just as would be expected if the excess were due to double pomeron vector boson pair production. (Again, pomeron exchange implies\cite{arw6} either a rapidity gap or, much more often, production of a high multiplicity state spread fairly uniformly across that part of the rapidity axis.) The vector boson decays would be via the (dominant) hadronic modes in this case, as opposed to the leptonic mode involved in the AMA-02 electroweak scale positron/electron excess shown in Fig.~1. 
 
Unfortunately, as is well known, more sophisticated jet algorithms and appropriately modified gluon distributions were searched for until the conclusion could be reached that there is no significant jet excess. Moreover, it has now become accepted that it is necessary/appropriate for large non-perturbative factors to be included in all applications of perturbative QCD at the LHC. As a result, ``non-perturbative'' hadronization corrections, that will 
surely obscure any large $E_T$ behavior due to sextet quark physics, are routinely included in jet cross-section analyses. 
\begin{center}

\epsfxsize=2.6in\epsffile{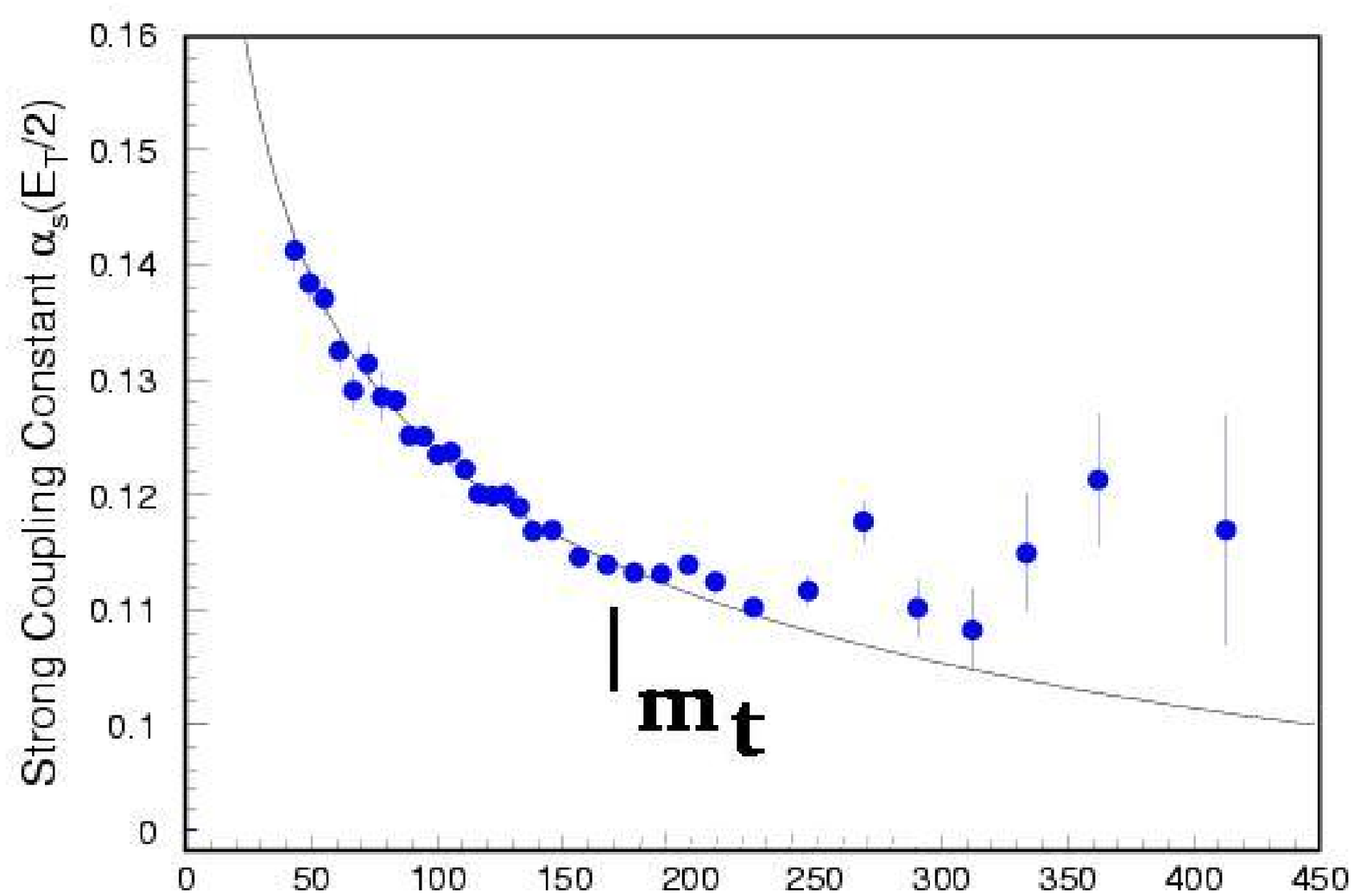} $~$ \epsfxsize=2.6in\epsffile{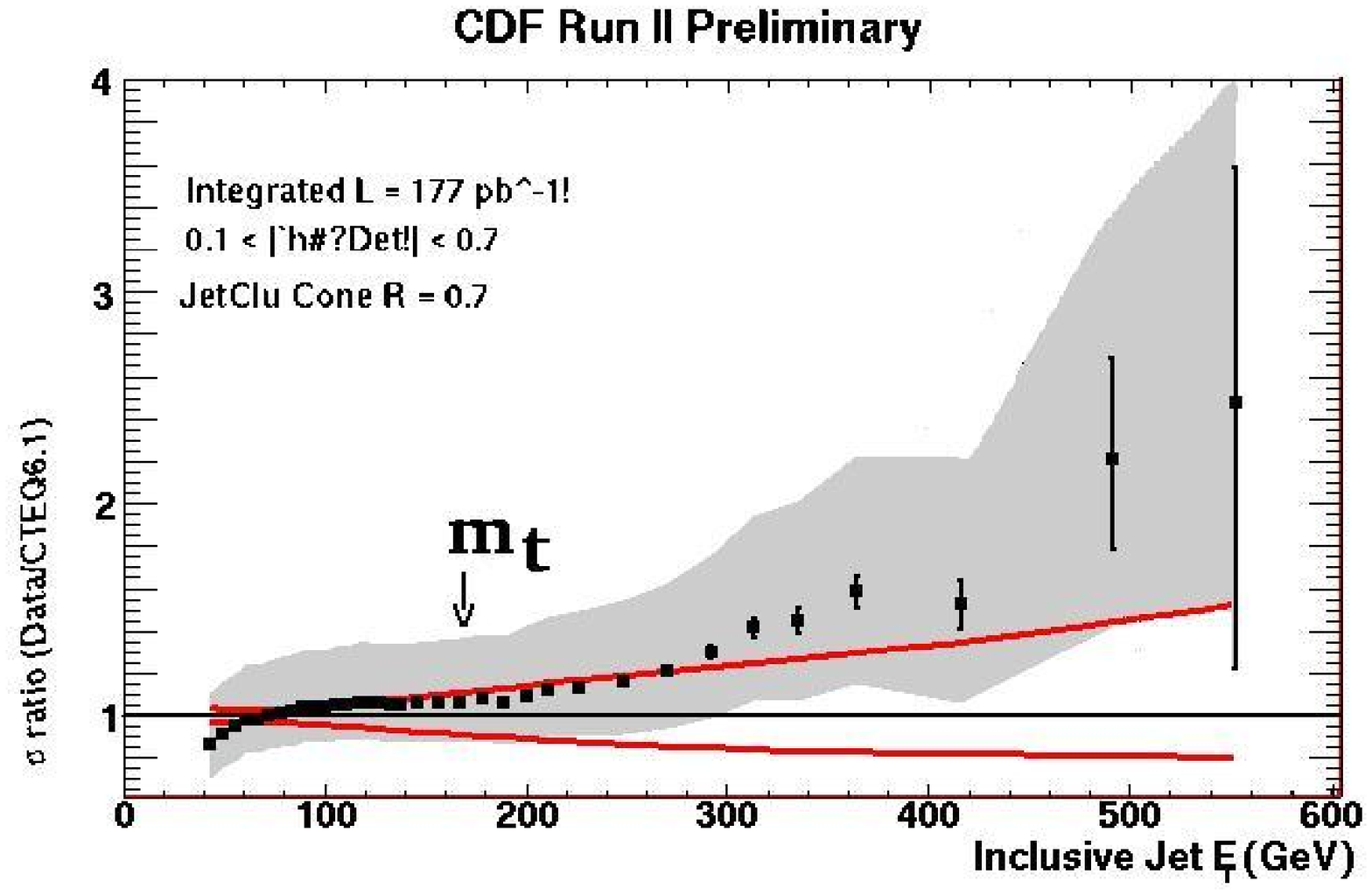}

(a) \hspace{2.7in} (b)

Figure 7. Early CDF data (a) $\alpha_S$ \cite{CDF}  - Run 1,  (b) Preliminary jet cross-section\cite{CDFe} -
Run 2.
\end{center}

As we have elaborated on in \cite{arw9}, because there is no derivation of the parton model, the current phenomenological determination of parton distributions from existing strong interaction data is consistent only if new short-distance physics is all that remains to be discovered. If new electroweak scale QCD physics is present, then the conventional parton model assumptions and procedure will clearly be inadequate.  
It is very fortunate, and may be of enormous historic importance, that the positrons 
recorded by AMS-02, unlike the jets seen by CDF, can not be made to disappear by
appealing to parton model ambiguities in the physics analysis and interpretation of the experimental results.
If the conclusion from the AMS-02 results is that the physics ``discovered'' in Fig.~7 is ``real'' then the implications for LHC physics are immeasurable.

That the (triplet quark) 
proton is lighter than the neutron is entirely due to the fact 
that the current mass of the $u$ quark is less than that of the $d$ quark. 
Electromagnetic effects, alone, 
would make the proton heavier. Remarkably, the current
quark mass difference produces an effect that is of the same order of magnitude,
but is just a little larger, and of the opposite sign, 
than the electromagnetic mass difference.

Because of the absence of hybrid triplet/sextet states, the lightest of the sextet nucleons must be stable. Also, sextet quark current masses must be zero if sextet pions 
are to mix, via anomaly poles, with the massless $W$ and $Z$ states to give 
them masses. In fact, in QUD, the SU(5) gauge bosons can directly transform triplet quarks to leptons, but can not do the same for sextet quarks. Consequently, massive triplet quarks, but not sextet quarks, would be generated (by self-energy corrections) in an effective lagrangian containing 
the massive physical leptons - obtained 
by ``integrating out'' the elementary massless leptons.

The sextet nucleon mass difference 
has to be entirely electromagnetic in origin, and so  
it is the neuson ($N_6$) and the antineuson ($\bar{N}_6$) that are stable. If the $\eta_6$ produces ``top events'', as discussed in the last Section, then the sextet
quark dynamical mass is approximately given by the Standard Model top quark mass. This implies the neuson mass should be  $\approx  500~ GeV$ and the proson
mass should be just a little higher. Since triplet and sextet quarks do not
combine to form bound states it can be assumed that
sextet nucleons also do not form bound states with triplet nucleons. More 
particularly, perhaps, if pion exchange provides the binding force for nucleons
to form nuclei, the distinct quark content of sextet and triplet nucleons
implies that there is no common ``pion'' that can bind them. 

The neuson is, therefore, neutral, stable, and (because of the dominance
of sextet states) will be the
dominant, heavy, stable state produced in high energy cross-sections. Consequently,
it will be dominantly produced in the high energy interactions that 
are believed to have been responsible for the formation of the early universe.
If it does not form bound states with normal quark matter it will 
abundantly form cold 
dark matter, in the form of (sextet) nuclei,  etc. (Perhaps
sextet pions can exist inside sextet nuclei and provide the binding force.) As 
a result, the existence of the sextet nucleon sector provides a natural
explanation for the dominance of dark matter in the universe. Conversely, if it can be established that neusons are responsible for dark matter, the dominance of dark matter
in the universe can be regarded as 
evidence confirming that sextet quark states dominate high energy cross-sections.
\begin{itemize}
\item{\bf QUD may provide a natural, and perhaps complete, explanation for the
existence of two distinct forms of matter in the universe, together with the distinct
mass scales and interaction properties.}
\end{itemize}
\begin{center}
 {\it This is all the more remarkable because, as we already emphasized,
QUD is self-contained, can not be added to, involves no parameters, and is entirely right, including the dark matter sector, or is simply wrong.}
\end{center}

\end{document}